\documentclass[10pt]{iopart}

\usepackage{graphicx}

\usepackage{verbatim}
\usepackage[dvips]{color}  
\usepackage{ulem} 

\begin{document}

\title[]{Magnetic field dependence of the precursor diamagnetism in La superconductors with magnetic Pr impurities}

\author{F Soto, C Carballeira, J M Doval, J Mosqueira, M V Ramallo, A Ramos-\'Alvarez, D S\'o\~nora, J C Verde, F Vidal}

\address{LBTS, Facultade de F\'isica, Universidade de Santiago de Compostela, ES-15782 Santiago de Compostela, Spain}


\begin{abstract}
The interplay between fluctuating Cooper pairs and magnetic impurities in conventional BCS low-$T_c$ superconductors has been studied through measurements of the magnetic field dependence of the fluctuation diamagnetism (FD) above $T_c$ in lanthanum with praseodymium impurities. These measurements provide a crucial confirmation of our previous observation [Europhys. Lett.~{\bf 73}~587 (2006)] that in the dilute impurity regime the FD increases almost linearly with the concentration of magnetic impurities. This striking effect is attributed to a variation due to the presence of the fluctuating Cooper pairs of the coupling between magnetic impurities. To describe these results at a phenomenological level, we propose a Gaussian Ginzburg-Landau model for the FD  which includes an indirect contribution proportional to both the impurities concentration and  the Cooper pairs density. Our approach is able to explain simultaneously the FD increase due to magnetic impurities and its decrease with the application of large magnetic fields.
\end{abstract}

\submitto{\SUST}
\maketitle

\ioptwocol

\section{Introduction}

The interplay between magnetism and superconductivity is still at present an open and interesting subject of strongly correlated electron systems \cite{science}. This thematic was opened by Matthias \textit{et al.} \cite{matias1,matias2} more than fifty years ago when they studied the decrease of the superconducting transition temperature ($T_c$) of lanthanum due to the introduction of magnetic rare earth impurities, an effect later explained in terms of pair breaking by Abrikosov and Gor'kov \cite{gorkov}. In spite of its interest, to our knowledge only a few works have addressed the interplay between magnetic impurities and the Cooper pairs created in the normal state by the thermal agitation (the so-called superconducting fluctuations \cite{tinkham,skocpolytinkham}), most of them dealing with the fluctuation-induced electrical conductivity (or paraconductivity), $\Delta\sigma$ \cite{Hobacuo,lindqvist,spahn}.

As early suggested on theoretical grounds by Schmidt \cite{newSchmidt}, magnetic impurities could enhance $\Delta\sigma$ through a change in the relaxation time of fluctuating Cooper pairs. Therefore, it could be expected that a time independent (static) observable like the fluctuation magnetic susceptibility $\Delta\chi$ will be unaffected by magnetic impurities, apart from the effects associated to possible changes in the superconducting parameters (for a recent theoretical work on this issue see, e.g., Ref.~\cite{borycki}). This is in fact the case of high-$T_c$ cuprates: recent measurements \cite{YBCO} have shown that the fluctuation diamagnetism of optimally-doped YBa$_2$Cu$_3$O$_{7-\delta}$ is not appreciably affected by the presence of magnetic impurities (both in the CuO$_2$ planes or intercalated between them). In striking contrast, the measurements in lanthanum with diluted magnetic (Pr) impurities presented in Ref.~\cite{epl} have shown an important enhancement of the fluctuation diamagnetism, that is observed to be well larger than the theoretical results  for three-dimensional isotropic superconductors \cite{tinkham,skocpolytinkham,schmidt-schmid,gollub1,gollub2,gollub3,PRLmosq,JPCMcarba}. The enhancement was found to be proportional to the concentration of impurities, being as large as a factor of 5 for a Pr concentration of 2 at.\%. This effect was attributed to a change in the coupling between magnetic ions (mediated by conduction electrons) due to the presence of the fluctuating Cooper pairs \cite{epl}. Nevertheless, those results were also affected by the experimental uncertainties associated with the strong temperature-dependence near $T_c$ (Curie-like) of the normal state background, and thus further experimental studies of these striking results are highly desirable. 

Here we present measurements of the magnetic-field-dependence of $\Delta\chi$ just above $T_c$ in La and La-Pr alloys. The used fields span from $5\times 10^{-3}$ T to 3 T, the latter value being three times larger than the upper critical field of the studied samples, where fluctuation effects are expected to be negligible \cite{skocpolytinkham,breakdown}. The interest of these new measurements is twofold: On the one side, the smooth $H$-dependence of $\chi$ in the normal state will allow a reliable determination of $\Delta\chi$ and to check the enhancement with the impurities concentration found in Ref.~\cite{epl}. On the other side, these new data will allow to study, for the first time in superconductors with magnetic impurities, the finite-field (or Prange) regime where fluctuation effects are expected to be strongly reduced with respect to the zero-field limit \cite{skocpolytinkham,breakdown}.

\section{Experimental details and results}

The samples used in this work (pure La and La-Pr alloys) are commercial (Alfa-Aesar and, respectively, Goodfellow; 99.9\% purity). They were used in the previous experiments of Ref.~\cite{epl}, and a thorough characterization may be found in Ref.~\cite{parameters}. The resulting superconducting parameters are summarized in Table~1. The $T_c$ values and the transition widths, $\Delta T_c$, were estimated from the temperature dependence of the low-field magnetic susceptibility. In the case of pure La, the transition temperature is well between the ones of pure $\alpha$-La \mbox{($\sim 5.0$ K)} and $\beta$-La \mbox{($\sim 6.0$ K)}  \cite{Leslie,Anderson,Berman,Finnemore,Johnson,Legvold-1977,Pan}, but no traces were observed of diamagnetic steps near these last temperatures. This is consistent with a mixing of both crystallographic phases at lengths of the order of or smaller than the superconducting coherence length amplitude $\xi(0)$ ($\sim20$~nm, see Table 1). The $\Delta T_c$ values are about 3\% the corresponding $T_c$ values, what will allow us to analyze fluctuation effects above $T_c$ down to reduced temperatures $\varepsilon\equiv\ln(T/T_c)\sim0.03$. In presence of a magnetic field, the $T_c(H)$ shift to lower temperatures further increases the accesible temperature range.
The Ginzburg-Landau (GL) parameter at $T_c$, $\kappa(T_c)$, and the upper critical field (linearly extrapolated to $T=0$~K), $H_{c2}(0)$, were obtained from magnetization measurements in the mixed state. In view of the $\kappa(T_c)$ values, even pure La is a type II superconductor, in agreement with earlier results of Pan and coworkers \cite{Pan}. We are not aware of previous $H_{c2}$ measurements in La-Pr alloys, but the results found in pure La are consistent with the ones in the literature. For instance, in Refs.~\cite{Anderson,Finnemore,Johnson,Pan} it is found a thermodynamic critical magnetic field $\mu_0H_C(0)\sim80-84$ mT for \mbox{$\alpha$-La} and $110-160$ mT for \mbox{$\beta$-La}. By using the relation $\mu_0H_C(0)=\mu_0H_{c2}(0)/\sqrt{2}\kappa$ and the data in Table~1, it results $\mu_0H_C(0)\approx95$ mT, well within these last values.
Finally, electron mean-free-paths at low-temperatures (estimated from the residual resistivities) are of the order or smaller than the corresponding $\xi(0)$, so that all the samples may be considered close to the dirty limit.\footnote{In the case of nominally pure La, this is consistent with the presence in the sample of a small amount of rare-earth impurities.} This excludes the presence of non-local effects on the fluctuation diamagnetism and, therefore, allows an easier comparison with the GL predictions.

\begin{table*}[t]
\begin{center}
\begin{tabular}{@{}lcccccc} 
\hline
Sample    &  $T_c$ & $\Delta T_c$ & $\mu_0H_{c2}(0)$ & $\xi(0)$ & $\kappa(T_c)$&$\ell$ \\ 
Pr at.\,\%&(K)&(K)&(T)&(nm)&&(nm) \\ 
\hline
0    &5.85&0.16&0.8&20&4.4&26.5 \\ 
0.5   &5.69&0.12&0.9&19&5.6&15.0 \\
1&5.51&0.25&0.9&18&6.0&11.0\\ 
2&5.40&0.18&1.0&18&6.4&7.5\\
\hline
\end{tabular}
\end{center}
\caption{\label{tab:Table}Summary of the superconducting parameters of the samples studied. The $T_c$ and $\Delta T_c$ values were obtained from the temperature dependence of the field-cooled magnetic susceptibility measured with $\mu_0H=0.5$~mT. $H_{c2}(0)$ and $\kappa$ were estimated from the reversible mixed-state magnetization (obtained as the average of the field-up and field-down measurements). The coherence length amplitudes were obtained as $\xi(0)=(\phi_0/2\pi\mu_0H_{c2}(0))^{1/2}$. The mean free paths were estimated from the resistivity just above $T_c$ by using a Drude model. For details see Ref.~\cite{parameters}.}
\end{table*}

%
%
\begin{figure}[t]
\begin{center}
\includegraphics[width=\columnwidth]{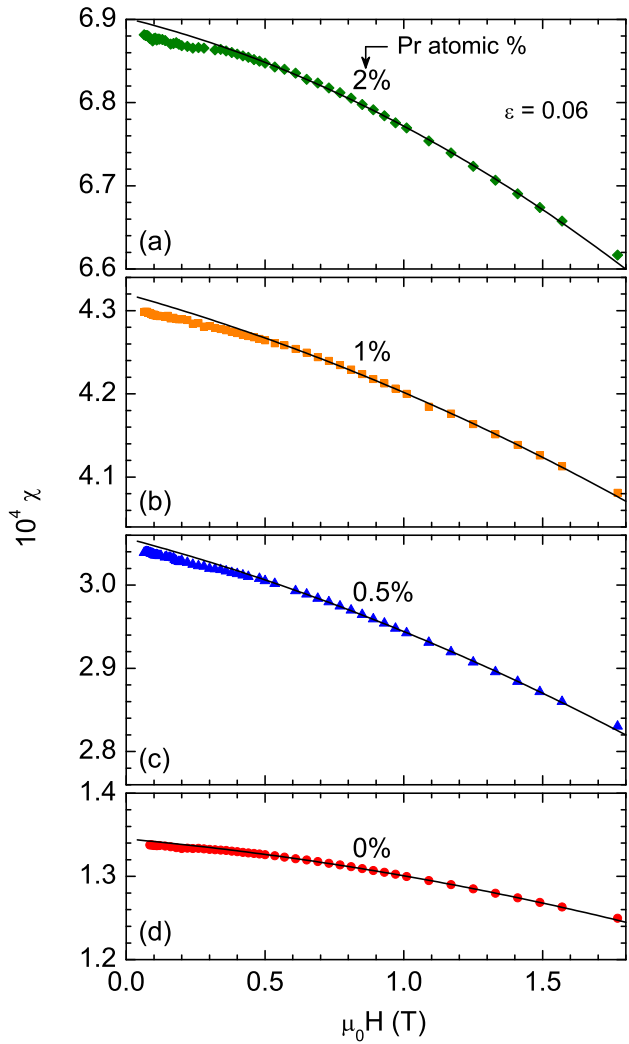}
\end{center} 
\caption{$H$-dependence of the magnetic susceptibility just above $T_c$ (at $\varepsilon=0.06$). The experimental uncertainty in $\chi$ is of the order of 0.01\%, and the corresponding error bars are not appreciable. The solid lines represent the normal-state susceptibility, determined by fitting a quadratic polynomial above $0.6H_{c2}(0)$, where fluctuation effects are expected to be negligible. These data are corrected for a $T$-independent upturn observed at low fields by subtracting the one observed at $\varepsilon=0.5$ (see Fig.~\ref{anomaly}).}
\label{backs}
\end{figure}

%
%
\begin{figure}[t]
\begin{center}
\includegraphics[width=\columnwidth]{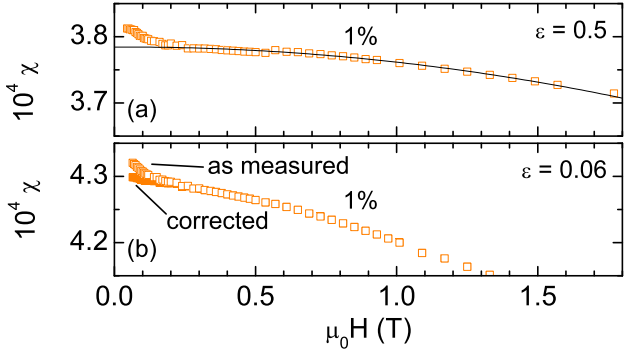}
\end{center} 
\caption{Example (for the sample with 1~at. \% Pr) of the procedure used to correct the $\chi(H)$ data for the $T$-independent upturn observed at low fields. The line in (a) is a fit of a quadratic polynomial to the $\chi(H)$ data taken at $\varepsilon=0.5$ (where fluctuation effects are negligible), in the $H$ region above the upturn. The resulting upturn is then removed from the data taken at $\varepsilon=0.06$, as it is shown in (b). The experimental uncertainty in $\chi$ is of the order of 0.01\%. For details see the main text.}
\label{anomaly}
\end{figure}

The magnetization measurements were performed with a commercial SQUID magnetometer (Quantum Design). In order to use all the available volume in the sample space, the samples were cut as cylinders of 6 mm in height and 6.5 mm in diameter. In Fig.~\ref{backs} we present the magnetic susceptibility of all the studied samples as a function of the  magnetic field at a selected reduced temperature, $\varepsilon\equiv\ln(T/T_c)=0.06$. Even La with 0\% Pr presents a paramagnetic behavior, probably due to the presence of a small amount ($\stackrel{<}{_\sim}0.1\%$) of magnetic rare-earth impurities. 
The data in Fig.~\ref{backs} are already corrected for a small upturn ($\stackrel{<}{_\sim}$~1\% of the normal state contribution) appearing typically below 0.15~T, see Fig.~\ref{anomaly}(a). This effect is of unknown origin,\footnote{It cannot be attributed to the paramagnetic contribution of the magnetic impurities, because it should be $H$-independent when $H\to0$. We also discarded that it is an artifact associated to a possible remnant magnetic field in the superconducting coil after setting it in persistent mode.} 
but it is unrelated to superconductivity because it is present in all isotherms up to well above $T_c$ ($\varepsilon=0.5$), where superconducting fluctuations are expected to be negligible \cite{vidal}. Therefore, we took advantage of its temperature independence and corrected the isotherms in the fluctuation region near $T_c$ by subtracting the anomaly observed at $\varepsilon=0.5$. An example of such a correction is shown in  Fig.~\ref{anomaly}. In any case this low-field anomaly has no consequence in our analysis, because it focuses in the high-$H$ region.

%
%
\begin{figure}[t]
\begin{center}
\includegraphics[width=\columnwidth]{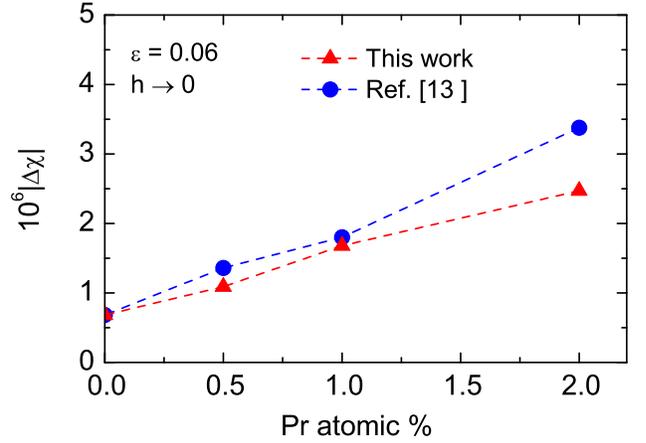}
\end{center} 
\caption{Dependence of the fluctuation magnetic susceptibility on the impurity concentration in the low field limit ($h<0.1$) and for $\varepsilon=0.06$. These data come from the $\chi(H)_T$ measurements in Fig.~\ref{backs}(a). For comparison, data from Ref.~\cite{epl} obtained from $\chi(T)_H$ measurements are also included. The experimental uncertainties in the $\Delta\chi$ values are about 2\%, and the corresponding error bars are smaller than the data points.}
\label{EfectovsX}
\end{figure}

The fluctuation-induced magnetic susceptibility $\Delta\chi$ may be obtained from the data in Fig.~\ref{backs} by subtraction of the (field-dependent) normal-state contribution. This last was estimated for each sample by fitting a quadratic polynomial in the interval $\sim(0.6-2)H_{c2}(0)$. The resulting background contributions are represented as solid lines in Fig.~\ref{backs}. The fitting function was chosen because it agrees with the data with a good accuracy (the maximum deviation in the fitting region is 0.5\%) and extrapolates smoothly to the lowest fields. Note that to get a better precision in determining the background contribution in the low-field region, we extended the lower bound of the fitting region down to 0.6$H_{c2}(0)$, neglecting the small fluctuation effects that may be present up to $\sim H_{c2}(0)$ \cite{breakdown}.

Fluctuation effects may be already appreciated in Fig.~\ref{backs} as a $\chi(H)$ reduction with respect to the background contribution for fields below $\sim0.5H_{c2}(0)$. The amplitude of this effect is roughly proportional to the Pr concentration, as shown in Fig.~\ref{EfectovsX}. As already commented in the Introduction, this striking effect was already observed in measurements of the fluctuation diamagnetism against temperature in the low-field limit \cite{epl}. The present measurements confirm those experimental findings, and will allow to study the $\Delta\chi$ behavior in presence of large reduced magnetic fields (in the so-called Prange fluctuation regime). 

\section{Data analysis}

The $\Delta\chi$ data will be analyzed in terms of a GL approach with Gaussian fluctuations of the superconducting order parameter (GGL approach) \cite{tinkham,skocpolytinkham}. The \textit{direct} contribution to $\Delta\chi$ (due to the Cooper pairs created by thermal fluctuations above $T_c$) was calculated in Ref.~\cite{JPCMcarba} (see also Refs.~\cite{PRLmosq,JPCMmosq}) by using a continuous Landau-level approximation (expected to be useful for reduced fields up to $h\sim0.3$) \cite{JPCMcarba}. This approach includes a total-energy cutoff to take into account the existence of a limit (of the order of the Cooper pairs size, $\xi_{T=0{\rm K}}$) to the shrinkage of the superconducting wavefunction at high reduced magnetic fields or temperatures. The resulting expression is
\begin{eqnarray}
\label{dmener}
\Delta \chi^{GGL}=-\frac{2\mu_0k_BT\xi(0)}{\phi_0^2h}\int_0^{\sqrt{c-\varepsilon}}dq\left[\frac{c-\varepsilon}{2h}\right.\nonumber\\
-\left.\left(\frac{c+q^2}{2h}\right)\psi\left(\frac{c+h+q^2}{2h}\right)+\ln\Gamma\left(\frac{c+h+q^2}{2h}\right)\right.\nonumber\\
+  \left.\left(\frac{\varepsilon+q^2}{2h}\right)\psi\left(\frac{\varepsilon+h+q^2}{2h}\right)-\ln\Gamma\left(\frac{\varepsilon+h+q^2}{2h}\right)\right],
\end{eqnarray}
where $\Gamma$ and $\psi$ are, respectively, the gamma and digamma functions, $k_B$ is the Boltzmann constant, $\phi_0$ the flux quantum, and $c\approx 0.55$ the cutoff constant \cite{vidal}. This expression includes as a particular case the low-field (or Schmidt) regime ($h\ll\varepsilon$), in which $\Delta\chi$ is field-independent: by imposing $h\ll \varepsilon$ in Eq.~(\ref{dmener}) we obtain 
\begin{eqnarray}
\label{smidE}
&&\Delta \chi^{GGL}=\nonumber\\
&&-\frac{\mu_0k_B\xi(0)T}{3\phi_0^2}\left(\frac{{\rm atan}\sqrt{\frac{c-\varepsilon}{\varepsilon}}}{\sqrt{\varepsilon}}-\frac{{\rm atan}{\sqrt{\frac{c-\varepsilon}{c}}}}{\sqrt{c}}\right)
\end{eqnarray}
that in absence of cutoff ($c\to\infty$) leads to the classical Schmidt result $\Delta\chi=-\pi\mu_0k_BT\xi(0)/6\phi_0^2\sqrt{\varepsilon}$ \cite{schmidt-schmid}.
Equations (\ref{dmener}) and (\ref{smidE}) predict the vanishing of $\Delta\chi$ at $\varepsilon=c$, the reduced temperature at which the GL coherence length $\xi(\varepsilon)$ becomes comparable to $\xi_{T=0{\rm K}}$ \cite{vidal}. 

%
%
\begin{figure}[t]
\begin{center}
\includegraphics[width=\columnwidth]{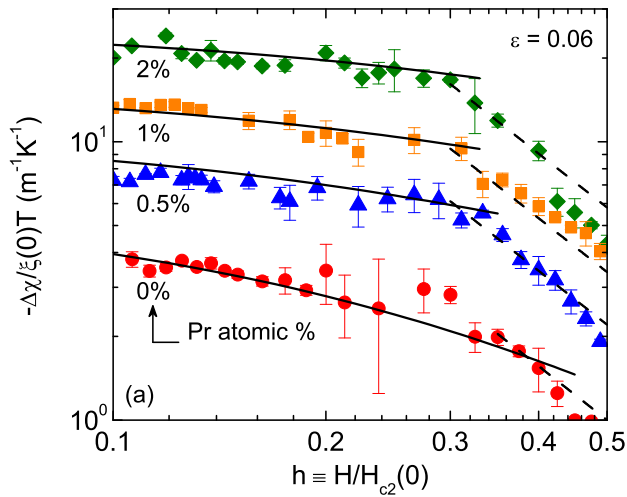}
\end{center} 
\caption{Reduced-field dependence of $\Delta\chi$ for $\varepsilon=0.06$ in all the studied samples. For an adequate comparison these data are normalized by $\xi(0)T$. The solid lines represent a fit of expression [\ref{final}], with $A\simeq-0.5$ as the only free parameter. The dashed lines are the empirical formula $(0.25+0.6\,x)\,h^{-2}$ (where $x$ is the Pr atomic \%) evidencing that the extra $\Delta\chi$ contribution due to Pr ions maintains its proportionality with  $x$ also for $h\stackrel{>}{_\sim}0.3$.}
\label{DX}
\end{figure}

The $h$-dependence of $\Delta\chi$ for all the studied Pr concentrations is shown in Fig.~\ref{DX}. To compare the results obtained in the different samples, in view of Eq.~(\ref{dmener}) these data are normalized by $\xi(0)T$. As expected \cite{PRLmosq,breakdown}, the experimental $\Delta \chi$ for pure La is in good agreement with Eq.~(\ref{dmener}). However, the $\Delta\chi$ amplitude increases with the Pr content (almost linearly, in view of Fig.~\ref{EfectovsX}). Moreover, on increasing the impurities concentration the $h$-dependence becomes less pronounced. These experimental evidences suggest that the magnetic impurities enhance some contribution, additional to $\Delta\chi^{GGL}$ having a smoother $h$-dependence.
As first suggested in Ref.~\cite{epl}, a natural candidate for the origin of such a contribution is the variation of the coupling between magnetic (Pr) impurities due to the condensation of Cooper pairs. 
The argument is as follows: in alloys with diluted magnetic impurities, the main vehicle for the interactions between these impurities is the electronic sea. A small variation of the density of normal electrons is expected to vary linearly the coupling strength between these impurities ({\it e.g.}, through the variation of the effective Fermi vector in the RKKY model of electron-mediated interaction between diluted magnetic impurities). As the condensation of Cooper pairs due to superconducting fluctuations will directly decrease the density of normal electrons, it may be expected that a contribution to the experimental $\Delta \chi$ will appear in the form
\begin{equation}
\label{adicional}
\Delta \chi^{\rm mag\ imp}= A\,x\,n_S(h)
\end{equation}
where $n_S(h)$ is the fluctuation-induced superfluid density, $A$ is an unknown constant, and $x$ is the magnetic impurity concentration (in atomic percents) to which $\Delta \chi^{\rm mag\ imp}$ is naturally expected to be proportional. Similar variations of the magnetization associated with the RKKY interaction due to the condensation of Cooper pairs and leading to effects proportional to $n_S$ were proposed previously, although not in the context of the superconducting fluctuations, by Anderson and Suhl for La-Gd systems in 1959 \cite{AndersonSuhl}, and more recently for certain borocarbides in Refs.~\cite{hede1,hede2}. 
Another reasoning proposed by Bergeret {\it et al.} \cite{Bergeret} based on magnetic moment screening (again not related with superconducting fluctuations) also lead to contributions to $\Delta \chi$ proportional to $xn_S$. 

%
%
\begin{figure}[t]
\begin{center}
\includegraphics[width=\columnwidth]{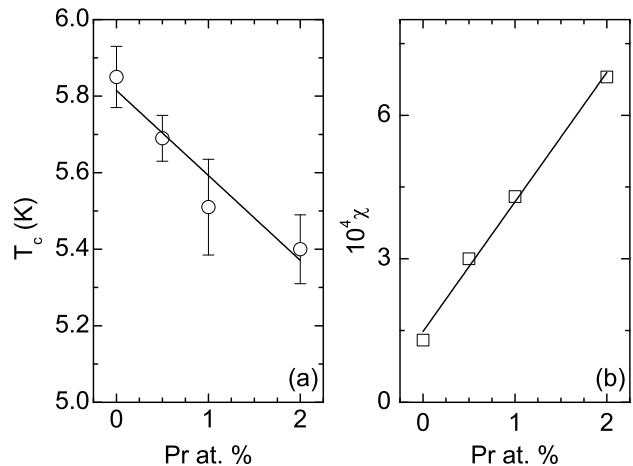}
\end{center} 
\caption{(a) $T_c$ dependence on the Pr content. The line is a fit of the Abrikosov-Gor'kov expression. (b) Linear increase of the magnetic susceptibility in the normal state (at a reduced temperature $\varepsilon=0.06$), as measured with $\mu_0H=1$~T. The experimental uncertainty in $\chi$ is of the order of 0.01\%, and thus the corresponding error bars are not appreciable.}
\label{magnetic}
\end{figure}

We calculate $n_S(\varepsilon,h)$ for an isotropic 3D superconductor on the same grounds as Eq.~(\ref{dmener}) (i.e., using the GGL approach above $T_c$ with a total-energy cutoff and for $h\stackrel{<}{_\sim}0.3$). For that, we apply to our $h\neq0$ case the same procedure as used in Ref.~\cite{vidal} for $h=0$ and substitute the $h=0$ fluctuation spectrum $\varepsilon+\xi^2(0)k^2$ by the $h>0$ one, $\varepsilon+(2n+1)h+\xi^2(0)k_z^2$. Here $k$ is the fluctuation wave vector, $k_z$ its component perpendicular to the applied magnetic field, and $n=0,1,\dots$ the Landau-level index. The GGL fluctuation-averaged superfluid density is then:
\begin{equation}
n_S=\left<\left|\Psi\right|^2\right>\propto\sum_{k_z,n}\frac{hT}{\varepsilon+(2n+1)h+\xi^2(0)k_z^2}
\label{suma}
\end{equation}
where $\left<\left|\Psi\right|^2\right>$ indicates the statistical and spatial average of the squared modulus of the GL wave function. Note the appearance in the $\sum_{k_z,n}$ summation of the $hT$ prefactor, that is due to the Boltzmann statistical weight and to the degeneration of each Landau level. The summation of Eq.~(\ref{suma}) becomes then possible using the same continuous Landau-level approach and total-energy cutoff procedure as used in Ref.~\cite{JPCMcarba} to obtain Eq.~(\ref{dmener}). The result is then (employing the same arbitrary dimensionless units as in Ref.~\cite{vidal}):
\begin{eqnarray}
n_S&=&\frac{\mu_0e^2k_BT\xi(0)}{2\hbar^2}\int_0^{\sqrt{c-\varepsilon}}dq\left[\psi\left(\frac{c+h+q^2}{2h}\right)\right.\nonumber\\
&-&\left.\psi\left(\frac{\varepsilon+h+q^2}{2h}\right)\right].
\end{eqnarray}
The solid lines in Fig.~\ref{DX} are the best fit of 
\begin{equation}
\Delta\chi^{GGL}(\varepsilon,h)+A\,x\,n_S(\varepsilon,h)
\label{final}
\end{equation}
to the $\Delta\chi(h)$ data for $\varepsilon=0.06$. The fitting region extends up to $h=0.3$, where the above approaches are expected to be applicable. As the data in this figure are already normalized by $\xi(0)T$, the only free parameter is $A$. The agreement is excellent, in some cases even above $h=0.3$. The resulting $A$ value ($-0.5$) is reasonably close to the one obtained in Ref.~\cite{epl} from low-field $\Delta\chi$ measurements against the temperature ($A\approx-0.7$), taking into account the important uncertainties associated to the background subtraction in this last case (mainly affecting the $\Delta\chi$ amplitude). 
The data in the region $h\stackrel{>}{_\sim}0.3$ can be fitted to the empirical formula $-\Delta\chi/\xi(0)T=(0.25+0.6\,x)\,h^{-2}$, indicating then that the extra $\Delta\chi$ contribution due to Pr ions maintains its proportionality with $x$ also for $h\stackrel{>}{_\sim}0.3$.

It is worth noting that Legvold \textit{et al.} \cite{Legvold} showed that the $T_c$ depression with the Pr concentration in La-Pr (which is consistent with the Abrikosov-Gor'kov theory, see Fig.~\ref{magnetic}(a)) is close to the one of non-magnetic La-Y and La-Lu alloys. This led them to suggest that Pr in the La-Pr alloy could be in a singlet state, an that there is little evidence for Pr-Pr interactions in these materials. In relation with this, in a recent paper Kogan and Prozorov \cite{Kogan} calculated that if the upper critical field increases with the impurities concentration, the system may not be in the purely magnetic scattering regime, and non-magnetic scattering should also be taken into account. However, our observed $H_{c2}(0)$ variation with the Pr concentration is quite similar to the experimental uncertainty.\footnote{See Ref.~\cite{parameters}, where the $H_{c2}(0)$ values were obtained from measurements of the reversible magnetization in the mixed state. Due to the highly irreversible nature of the latter in La-Pr alloys, $H_{c2}(0)$ was obtained as the average of the field-up and field-down contributions. The uncertainty associated with this procedure is comparable with the seeming variation of $H_{c2}(0)$ with $x$.} By this reason, in our simplified model we have ignored a possible contribution coming from non-magnetic scattering. Also, the linear increase of the magnetic susceptibility in the normal state with the Pr concentration (see Fig.~\ref{magnetic}(b)) clearly indicates that Pr does introduce some moderate additional magnetic moment in the alloy. Our present results indicate thus that even these relatively weak magnetic moments (even if perhaps they do not dominate the $T_c(x)$ shift) may profoundly affect the diamagnetic contribution coming from superconducting fluctuations above $T_c$.

The present results contrast with the behavior observed in high-$T_c$ cuprates, where the magnetic impurities (within the CuO$_2$ layers or intercalated between them) lead to inappreciable effects in the fluctuation-induced magnetic susceptibility once the change in the $T_c$ value is taken into account. This suggests that in these materials the pairs created above $T_c$ by thermal fluctuations lead to a negligible effect on the interaction between magnetic impurities.

\section{Conclusions} 

We have presented detailed measurements of the diamagnetism induced by superconducting fluctuations in a BCS superconductor (La) under the simultaneous presence of magnetic impurities (Pr) and external magnetic fields. It is found that magnetic fields approaching the upper critical field at $T\to0$~K are very effective reducing the fluctuation effects. However, counter intuitively at first, the magnetic impurities lead to an important enhancement of the amplitude of the fluctuation magnetic susceptibility. This enhancement is roughly proportional to the concentration of magnetic impurities, and is as large as one order of magnitude for concentrations of only 2 at.~\% of Pr in La. 

The experimental data are interpreted in terms of a change of the coupling between magnetic impurities (mediated by conduction electrons), induced by the presence of fluctuating Cooper pairs. According to this, to analyze the experimental data we proposed a Gaussian Ginzburg-Landau model for $\Delta\chi$ which includes an \textit{indirect} contribution proportional to the impurities concentration, $x$, and to the Cooper pairs density, $n_S$. This approach includes a total-energy cutoff to account for the short-wavelength modes excited at high reduced magnetic fields, and is expected to be applicable up to $\sim0.3H_{c2}(0)$. The agreement with the experimental data is excellent in a wide reduced magnetic field interval and for all the  studied impurities concentration, in agreement with previous findings in the low-field limit \cite{epl}. It would be interesting to further investigate whether fully microscopic approaches may account for the amplitude of the impurity-induced enhancement of the fluctuation diamagnetism.

\ack

This work was supported by the Spanish MICINN (grant No.~FIS2010-19807), and by the Xunta de Galicia (grant No.~GPC2014/038). A.\,R-A. acknowledges support through a FPI grant.

\end{document}